\def\be{\begin{equation}}
\def\ee{\end{equation}}
\def\ba{\begin{array}}
\def\ea{\end{array}}

\def\Rb{{I\!\! R}}

\def\Cb{\ \hbox{\vrule width 0.6pt height 6pt depth 0pt
              \hskip -3.2 pt} C}
\newtheorem{definition}{Definition}
\newtheorem{theorem}{Theorem}

\newtheorem{proposition}{Proposition}
\newtheorem{corollary}{Corollary}

\documentstyle[12pt]{article}
\begin{document}
\parskip=4pt
\parindent=18pt
\baselineskip=22pt
\setcounter{page}{1}
\centerline{\Large\bf
Point Interactions:  ${\mathcal P \mathcal T}$-Hermiticity}
\vspace{2ex} \centerline{\Large\bf and Reality of the Spectrum}
\vspace{3ex}

\centerline{S. Albeverio$^{1}$, S-M. Fei$^{2}$ and P.
Kurasov$^3$}

\parskip=0pt
\parindent=0pt
\baselineskip=16pt \vskip 1 true cm

\begin{center}
\begin{minipage}{5.3in}
$^1$~Institute f\"ur Angewandte Mathematik, Univ. Bonn, 53155 Bonn

\parindent=9pt
SFB 256 Bonn, BiBoS, Bielefeld-Bonn, CERFIM, Locarno and USI

albeverio@uni-bonn.de.

\parindent=0pt
$^2$~Institute f\"ur Angewandte Mathematik, Univ. Bonn, 53155 Bonn

\parindent=9pt
Dept. of Mathematics, Capital Normal University, Beijing 100037

fei@uni-bonn.de

\parindent=0pt
$^3$~Dept. of Mathematics, Lund Institute of Technology, Box 118,
221 00

\parindent=9pt
Lund, SWEDEN, kurasov@maths.lth.se.

\end{minipage}
\end{center}

\parindent=18pt
\parskip=6pt

\begin{center}
\begin{minipage}{5in}
\vspace{3ex} \centerline{\large Abstract} \vspace{4ex} General
point interactions for the second derivative operator in one
dimension are studied. In particular, ${\mathcal P \mathcal
T}$-self-adjoint point interactions with the support at the origin
and at points $\pm l$ are considered. The spectrum of such
non-Hermitian operators is investigated and conditions when the
spectrum is pure real are presented. The results are compared with
those for standard self-adjoint point interactions.
\end{minipage}
\end{center}
\vspace{5ex} \baselineskip=18pt



\section{Introduction.}

Exactly solvable models are used in quantum mechanics to obtain
Hamiltonians describing realistic physical systems but having the
important property of being exactly solvable, i.e. that all
eigenfunctions, spectrum and scattering matrix can be calculated
in closed form using elementary functions. A large class of such
operators can be obtained using the method of point interactions
described in detail in \cite{aghkh} in application to the theory
of self-adjoint operators. It is not surprising that point
interactions can be used to obtain exactly solvable non
self-adjoint operators, see for example \cite{pavlov-proc,pavlov},
where the Schr\"odinger operator on the half-axis with complex
boundary condition at the origin has been considered as a model
for dissipative operators. Using dissipative operators instead of
self-adjoint ones one obtains Hamiltonians describing irreversible
quantum systems which are of great importance especially for
evolution problems.
 It has been discovered recently that Hamiltonians
possessing so-called $ {\mathcal P \mathcal T }$-symmetry can be
non self-adjoint and have the real spectrum at the same time. This
discovery observed first using numerical computations has been
approved by extensive analytical calculations. Many Schr\"odinger
operators which have been studied have complex $ {\mathcal P
\mathcal  T}$-symmetric potentials but real discrete spectrum.
Even if no proof that the spectrum is real for the whole class of
operators has been discovered, it was expected that the relations
between these two properties are rather close. In the present
article we consider ${\mathcal P \mathcal T}$-symmetric operators
with point interactions, which are exactly solvable. This gives us
an opportunity to study the relations between the $ {\mathcal P
\mathcal  T}$-symmetry of the Hamiltonian and the reality of its
spectrum in full detail. We show in particular that none of these
two properties imply the other one, but that there exist
Hamiltonians with point interactions which are ${\mathcal
P\mathcal T}$-symmetric and have real spectrum. Since the
operators under investigation are not self-adjoint the classical
von Neumann theory cannot be applied without modification. Instead
the method of boundary conditions is used. It is shown how the
notion of $ {\mathcal P \mathcal T}$-symmetry has to be modified
in application to the extension theory for linear operators. In
addition we introduce the notion of $ {\mathcal P \mathcal
T}$-self-adjoint operators.
 The operators obtained can be used to describe irreversible
systems in the framework of the recently appeared $ {\mathcal P
\mathcal T}$-symmetric quantum mechanics described below.

Let us start by presenting recent results concerning
one-dimensional Schr\"odinger operator with $ {\mathcal P \mathcal
T}$-symmetric potentials determined by the following expression

\begin{equation} \label{def} L = - \frac{d^2}{dx^2} + V(x),
\end{equation}
where the potential $ V(x)$ is not a real valued function, but
satisfies:
$$ V(-x) = \overline{V(x)} .$$
Using the space parity operator $ {\mathcal P}: ({\mathcal P}
\psi) (x) = \psi (-x)$ and the complex conjugation operator $
{\mathcal T} \psi = \overline{\psi}$ the last property can be
written as
$$ {\mathcal P \mathcal T} V = V.$$
Then the differential operator $ L $ is formally $ {\mathcal P
\mathcal T} $-symmetric
$$ {\mathcal P \mathcal T} \; L = L \;
 {\mathcal P \mathcal T}.$$
Interest to differential operators with such symmetry became
enormous, after it has been discovered that some of these
operators have real spectrum (like self-adjoint operators) and
therefore can be used in construction of a new, $ {\mathcal P
\mathcal T}$-symmetric quantum mechanics.\footnote{It is not true
in general that the spectrum of every $\mathcal P \mathcal
T$-symmetric operator is real.}
 This fact was first
discovered numerically by C.M.Bender, S.Boettcher and
collaborators
\cite{bender5,bender2,bender1,bender3,bender4,bender6}. Analytical
studies of $ {\mathcal P \mathcal T}$-symmetric Hamiltonians were
carried out by M.Znojil and his collaborators
\cite{zn2,zn7,zn8,zn9,zn3,zn4,zn6,zn5,zn1}. Numerous
non-self-adjoint operators with real spectrum were studied
 (see e.g. \cite{dorey,dorey2,handy,handy2}).

The aim of the current paper is to describe exactly solvable $
\mathcal P \mathcal T $-Hermitian operators constructing using the
method of point interactions \cite{aghkh,book}. Standard
symmetries of (self-adjoint) point interactions in dimension $ 1$
were studied in \cite{dabr}. We are going now to extend these
results to include ${\mathcal P \mathcal T}$-self-adjoint point
interactions. Starting from the second derivative operator in $
L_2 (\Rb) $ (which is $\mathcal P \mathcal T$-self-adjoint) we
construct non trivial point interactions leading to operators
having $ \mathcal P \mathcal T$-symmetry. Their spectral
properties are described in Section 3. Section 4 is devoted to
construction of local point interactions concentrated at different
points on the real line. One of the main questions investigated is
the relations between the ${\mathcal P \mathcal T}$-symmetry of
the operators and reality of their spectrum. The family of point
interactions with real spectrum is characterized in Section 5.

\section{ ${\mathcal P \mathcal T}$-self-adjoint point interactions.}

The method of point interactions is well described in several
monographs \cite{aghkh,book,demkov}. This method is based on the
fact that differential expression which is formally symmetric (or
$ {\mathcal P \mathcal T}$-symmetric in our case) does not
determine the operator in the Hilbert space uniquely. To determine
the operator one has to specify its domain. One can start from a
certain standard differential operator (i.e. Laplace operator) and
restrict it to a certain densely defined operator. Then extending
the restricted operator to another ${\mathcal P \mathcal
T}$-Hermitian operator one can get an operator with nontrivial
spectral structure.
 Thus the main tool to be used is the extension
theory for linear operators. The method we are going to present is
similar to the method of point interaction developed for
self-adjoint operators. Therefore let us introduce two definitions
which are similar to the definition of symmetric and self-adjoint
operators in conventional quantum mechanics.

\begin{definition}
An operator $ L $  is called {\bf $ {\mathcal P \mathcal
T}$-symmetric}\footnote{The operators described by these
definitions should be better called $ {\mathcal P}$-symmetric and
${\mathcal P}$-self-adjoint, but we prefer to use notation
${\mathcal P \mathcal T}$-symmetric and ${\mathcal P \mathcal
T}$-self-adjoint in order to underline the entire relations with $
{\mathcal P \mathcal T}$-symmetric quantum mechanics already
described in the literature.} if it is densely defined and the
following inclusion holds
\begin{equation} \label{ptsym}
L^* \supset {\mathcal P} L {\mathcal P} .
\end{equation}
A ${\mathcal P \mathcal T}$-symmetric operator $ L $ is called
{\bf ${\mathcal P \mathcal T}$-self-adjoint} if and only if
\begin{equation} \label{ptsa}
L^* ={\mathcal P} L {\mathcal P} .
\end{equation}
\end{definition}

The second derivative operator $ L = - \frac{d^2}{dx^2} $ with the
standard domain $ W_2^2 ({\Rb}) $ is both self-adjoint and
${\mathcal P \mathcal T}$-self-adjoint. In the current section we
study point perturbations at the origin of this operator leading
to ${\mathcal P \mathcal T}$-self-adjoint operators. By  point
interaction we mean any interaction which vanishes on the
functions with the support separated from the origin. In other
words a linear operator $ A $ is a point perturbation at the
origin of the operator $ L $ if and only if its restriction to $
C_0^\infty ({\Rb} \setminus \{ 0 \}) $ coincides with the
restriction of the operator $ L $:
$$ A \vert_{C_0^\infty ({\Rb} \setminus \{ 0 \}) } =
L \vert_{C_0^\infty ({\Rb} \setminus \{ 0 \}) } \equiv L_0 . $$
The operator $ L_0 $ determined by the last equation is both
symmetric and ${\mathcal P \mathcal T}$-symmetric. Moreover
$$ {\mathcal P} L_0 {\mathcal P} = L_0, \; \; {\mathcal P} L_0^* {\mathcal P} = L_0^*, $$
and thus any $ {\mathcal P \mathcal T}$-symmetric extension of $
L_0 $ is a restriction of the maximal operator $ L_{\rm max} =
L_0^* $ being the second derivative operator in $ L_2 ({\Rb}) $
with the domain $ {\rm Dom} (L_{\rm max}) = W_2^2 ({\Rb} \setminus
\{ 0 \} ). $ All such $\mathcal P \mathcal T$-self-adjoint
extensions and therefore all $ {\mathcal P \mathcal
T}$-self-adjoint point interactions at the origin are described by
the following

\begin{theorem}
The family of ${\mathcal P \mathcal T}$-self-adjoint second
derivative operators with  point interactions at the origin
coincides with the set of restrictions of the second derivative
operator $ L_{\rm max} = - \frac{d^2}{dx^2}$, defined originally
on $ W_2^2 ({\Rb} \setminus \{0\})$, to the domain of functions
satisfying the boundary conditions at the origin of one of the
following two types
\newline
$ {\rm I.} $\begin{equation} \label{bcond1} \left(\begin{array}{c}
\psi (+0) \\
\psi' (+0)
\end{array}\right) = B
 \left(\begin{array}{c}
\psi (-0) \\
\psi' (-0)
\end{array}\right);
\end{equation}
with the matrix $ B$ equal to
$$ B = e^{i \theta}
\left( \begin{array}{cc}
\sqrt{1 +bc} \; e^{i\phi} & b \\
c & \sqrt{1+bc} \; e^{-i\phi}
\end{array} \right) $$
with the real parameters $ b \geq 0, c \geq -1/b,$\footnote{If the
parameter $b$ is equal to zero, then the last inequality can be
neglected.} $ \theta, \phi \in [0, 2 \pi);$
\newline
$ {\rm II.}$
\begin{equation} \label{bcond2}
\left\{
\begin{array}{ccc}
h_0 \psi' (+0) & = & h_1 e^{i\theta} \psi (+0)\\
h_0 \psi' (+0) & = & - h_1 e^{-i\theta} \psi (-0)
\end{array}
\right.
\end{equation}
with the real phase parameter $ \theta \in [0,2\pi)$ and with the
parameter $ {\bf h} = (h_0, h_1)$ taken from the (real) projective
space $ {\bf P}^1.$
\end{theorem}
{\bf Proof}
 We have already proven that any $ {\mathcal P \mathcal
T}$-self-adjoint extension of $L_0$ is a restriction of $ L_{\rm
max}.$ Consider the vector space $ {\Cb}^4 $ of boundary values of
functions from the domain of $ L_{\rm max} $ and the map $ \Gamma$
adjusting to any  function $ \psi \in W_2^2 ({\Rb} \setminus \{ 0
\})$ its boundary values
$$ \Gamma: \psi \mapsto \left(\begin{array}{c}
\psi (+0)\\
\psi' (+0) \\
\psi (-0) \\
\psi' (-0)
\end{array}\right) \in {\Cb}^4.$$
The fundamental symmetry operator $ {\mathcal P}$ acting in the
space of boundary values coincides with the operator of
multiplication by the following matrix $ P$
\begin{equation}
P \Gamma = \Gamma {\mathcal P}, \;\;\; {\rm where} \; \; \; P =
\left(
\begin{array}{cccc}
0 & 0 & 1 & 0 \\
0 & 0 & 0 & -1 \\
1 & 0 & 0 & 0 \\
0 & -1 & 0 & 0
\end{array}
\right).
\end{equation}
Similarly the operator of complex conjugation $ {\mathcal T}$ is
mapped by $ \Gamma$ into the operator $T$ of complex conjugation
in $ {\Cb}^4.$

The closure of the operator $ L_0 $ is defined on the functions
having trivial boundary values at the origin and {\it Vice Versa}
any function from the domain of $ L_{\rm max} $ with trivial
boundary values belongs to $ {\rm Dom} (\overline{L_0}). $ Thus
the dimension of the quotient space $ {\rm Dom} \,(L_{\rm max}) /
{\rm Dom}\, (\overline{L_0})$ is equal to $ 4$.  Any $ {\mathcal P
\mathcal T}$-self-adjoint extension of $ L_0$ can be described as
the restriction of $ L_{\rm max}$ to the set of functions with
boundary values from a certain $ 2$-dimensional subspace
${\mathcal L}$ of $ {\Cb}^4$. Every such subspace can be described
by
 two (linearly independent)
boundary conditions using a certain $ 2 \times 4$ dimensional rank
$ 2$ matrix $Q = \{ q_{ij}\}$ as follows
\begin{equation}
\label{del} \Gamma \psi \in {\mathcal L} \Leftrightarrow \left\{
\begin{array}{ccc}
q_{11} \psi (+0) + q_{12} \psi'(+0) + q_{13} \psi (-0) + q_{14}
\psi'(-0)
& = & 0 \\
q_{21} \psi (+0) + q_{22} \psi'(+0) + q_{23} \psi (-0) + q_{24}
\psi'(-0) & = & 0
\end{array}
\right. .
\end{equation}
The restriction of the maximal operator to a two-dimensional
subspace $ {\mathcal L}$ possesses the $ {\mathcal P \mathcal
T}$-symmetry if and only if
$$ \vec{X} \in {\mathcal L} \Leftrightarrow  PT \vec{X} \in {\mathcal L}.$$

Let us prove now that the $ {\mathcal P \mathcal T}$-self-adjoint
point interactions can be parameterized in at least one of the
following four ways
\newline $ {\bf A} $
\begin{equation} \label{bc1}
\left(\begin{array}{c}
\psi (+0) \\
\psi' (+0)
\end{array}\right) =
e^{i \theta} \left( \begin{array}{cc}
\sqrt{1 +bc} \; e^{i\phi} & b \\
c & \sqrt{1+bc} \; e^{-i\phi}
\end{array} \right) \left(\begin{array}{c}
\psi (-0) \\
\psi' (-0)
\end{array}\right);
\end{equation}
$ b \geq 0, c \geq -1/b, \theta, \phi \in [0, 2 \pi);$
\newline $ {\bf B}$
\begin{equation} \label{bc3}
\left(\begin{array}{c}
\psi (+0) \\
\psi' (-0)
\end{array}\right) =
e^{i \theta} \left( \begin{array}{cc}
\sqrt{1 +bc} \; e^{i\phi} & b \\
c & \sqrt{1+bc} \; e^{-i\phi}
\end{array} \right) \left(\begin{array}{c}
\psi (-0) \\
\psi' (+0)
\end{array}\right),
\end{equation}
$ b \geq 0, c \geq -1/b, \theta, \phi \in [0, 2 \pi);$
\newline ${\bf C}$
\begin{equation} \label{bc2}
\left(\begin{array}{c}
\psi (+0) \\
\psi (-0)
\end{array}\right) =
\left(
\begin{array}{cc}
a e^{i\theta} & b e^{i\phi} \\
- b e^{-i\phi} & - a e^{-i\theta}
\end{array}
\right)
 \left(\begin{array}{c}
\psi' (+0) \\
\psi' (-0)
\end{array}\right),
\end{equation}
$ a, b \geq 0, \theta, \phi \in[0, 2\pi);$
\newline
${\bf D}$
\begin{equation} \label{bc5}
\left(\begin{array}{c}
\psi' (+0) \\
\psi' (-0)
\end{array}\right) =
\left(
\begin{array}{cc}
a e^{i\theta} & b e^{i\phi} \\
- b e^{-i\phi} & - a e^{-i\theta}
\end{array}
\right) \left(\begin{array}{c}
\psi (+0) \\
\psi (-0)
\end{array}\right),
\end{equation}
$ a, b \geq 0, \theta, \phi \in[0, 2\pi).$

Since the matrix $ Q$ appearing in (\ref{del}) has rank $2$, at
least one of its six  $ 2 \times 2$ minors is non-degenerate.
Depending on which minor is non-degenerate the boundary conditions
(\ref{del}) can be written in different ways using certain $ 2
\times 2$ matrices $ B = \left( \begin{array}{cc}
\alpha & \beta \\
\gamma & \delta
\end{array} \right)$
\newline
1) $ \det \left(
\begin{array}{cc}
q_{11} & q_{12} \\
q_{21} & q_{22}
\end{array} \right) \neq 0
\Rightarrow  \left(\begin{array}{c}
\psi (+0) \\
\psi' (+0)
\end{array}\right) =
B \left(\begin{array}{c}
\psi (-0) \\
\psi' (-0)
\end{array}\right).
$
\newline
2) $ \det \left(
\begin{array}{cc}
q_{11} & q_{13} \\
q_{21} & q_{23}
\end{array} \right) \neq 0
\Rightarrow  \left(\begin{array}{c}
\psi (+0) \\
\psi (-0)
\end{array}\right) =
B \left(\begin{array}{c}
\psi' (+0) \\
\psi' (-0)
\end{array}\right).
$
\newline
3)  $ \det \left(
\begin{array}{cc}
q_{11} & q_{14} \\
q_{21} & q_{24}
\end{array} \right) \neq 0 \Rightarrow
 \left(\begin{array}{c}
\psi (+0) \\
\psi' (-0)
\end{array}\right) =
B \left(\begin{array}{c}
\psi (-0) \\
\psi' (-0)
\end{array}\right).
$
\newline
4)  $ \det \left(
\begin{array}{cc}
q_{12} & q_{13} \\
q_{22} & q_{23}
\end{array} \right)\neq 0 \Rightarrow
 \left(\begin{array}{c}
\psi (-0) \\
\psi' (+0)
\end{array}\right) =
B \left(\begin{array}{c}
\psi (+0) \\
\psi' (-0)
\end{array}\right).
$
\newline
5)  $  \det \left(
\begin{array}{cc}
q_{12} & q_{14} \\
q_{22} & q_{24}
\end{array} \right) \neq 0 \Rightarrow
 \left(\begin{array}{c}
\psi' (+0) \\
\psi' (-0)
\end{array}\right) =
B \left(\begin{array}{c}
\psi (+0) \\
\psi (-0)
\end{array}\right).
$
\newline
6)  $ \det \left(
\begin{array}{cc}
q_{13} & q_{14} \\
q_{23} & q_{24}
\end{array} \right) \neq 0 \Rightarrow
 \left(\begin{array}{c}
\psi (-0) \\
\psi' (-0)
\end{array}\right) =
B \left(\begin{array}{c}
\psi (+0) \\
\psi' (+0)
\end{array}\right).
$
\newline
Our aim now is to characterize all matrices $ B$ leading to $
{\mathcal P \mathcal T}$-symmetric boundary conditions.

Consider the first case. Suppose that the function $ \psi$
satisfies the boundary conditions. The boundary conditions for the
function $ {\mathcal P \mathcal T} \psi$ are given by
$$ \left(
\begin{array}{c}
\psi(-0) \\
\psi' (-0)
\end{array}
\right) = \left( \begin{array}{cc}
- \bar{\alpha} & - \bar{\beta} \\
- \bar{\gamma} & - \bar{\delta}
\end{array}\right) \left(
\begin{array}{c}
\psi(+0) \\
\psi' (+0)
\end{array}
\right).$$ These conditions coincide with the original one if and
only if
\begin{equation} \label{sl1} \left( \begin{array}{cc}
- \bar{\alpha} & - \bar{\beta} \\
- \bar{\gamma} & \bar{\delta}
\end{array}\right) =  \left( \begin{array}{cc}
\alpha & \beta \\
\gamma & \delta
\end{array} \right)^{-1}.
\end{equation}
Investigating cases 3), 4) and 6) one arrives to exactly the same
equation on the matrix $ B.$

Equation (\ref{sl1}) implies that $ \det B = \alpha \delta - \beta
\gamma \neq 0$ and the following four equalities hold
$$
\begin{array}{ccccccc}
\bar{\alpha} & = \frac{1}{\det B} \delta , & &
\bar{\beta} & = \frac{1}{\det B} \beta , \\
&&&&&& \\
 \bar{\gamma} & = \frac{1}{\det B} \gamma, & &
\bar{\delta} & = \frac{1}{\det B} \alpha .\\
\end{array}
$$ These equations imply in particular that the
determinant has absolute value $ 1,$ i.e. there exists $ \theta
\in [0, 2\pi)$ such that $ \det B = e^{2i\theta}.$ Consider then
the matrix
$$ B' = \left( \begin{array}{cc}
\alpha' & \beta' \\
\gamma' & \delta'
\end{array} \right) = e^{-i \theta} B ,$$
with the unit determinant $ \alpha' \delta' - \beta' \gamma' = 1$.
The entries of $ B'$ satisfy the following equations
$$
\begin{array}{ccccccc}
\bar{\alpha}' & = & \delta' ,&&
\bar{\beta}' & = & \beta' ,\\
&&&&&& \\
\bar{\gamma}' & = & \gamma', &&
\bar{\delta}' & = & \alpha' .\\
\end{array}
 $$ Hence all such matrices $ B'$ can be parameterized by $ 3$
real parameters:
\newline \indent positive number $b$;
\newline  \indent real number $c, c \geq -1/b$;
\newline \indent phase parameter $ \phi \in [0, 2 \pi);$
\newline
using the following formula
\begin{equation}
B' = \left( \begin{array}{cc}
\sqrt{1+bc}\; e^{i\phi} & b \\
c & \sqrt{1+bc} \; e^{-i\phi}
\end{array} \right).
\end{equation}
This implies that the boundary conditions can be written in the
form (\ref{bc1}) in the cases 1) and 6) and in the form
(\ref{bc3}) in the cases 3) and 4).\footnote{One has to take into
account that $ \det B \neq 0$ and therefore the boundary
conditions 1) and 6) and 3) and 4) are pairwise equivalent.}

It remains to study the cases 2) and 5). The boundary conditions
determine a $ {\mathcal P \mathcal T}$-symmetric operator if and
only if the following equality holds in both cases
\begin{equation}
\left( \begin{array}{cc}
\alpha & \beta \\
\gamma & \delta
\end{array} \right) = -
\left( \begin{array}{cc}
\bar{\delta} & \bar{\gamma} \\
\bar{\beta} & \bar{\alpha}
\end{array} \right)
\end{equation}
This equality is satisfied if and only if
$$ \left\{
\begin{array}{ccc}
\alpha & = & - \bar{\delta} \\
\beta & = & - \bar{\gamma}
\end{array}
\right. ,$$ and such matrices can be parameterized by the
following parameters:
\newline \indent two positive numbers $ a, b \in {\Rb}_+;$
\newline \indent two phases $ \theta, \phi \in [0, 2 \pi);$
\newline
using the formula
$$ B = \left(
\begin{array}{cc}
a e^{i\theta} & b e^{i\phi} \\
- b e^{-i\phi} & - a e^{-i\theta}
\end{array}
\right).$$ We thus get the parameterizations (\ref{bc2}) and
(\ref{bc5}).

It remains to prove that each of the boundary conditions $ {\bf
B}, {\bf C}$ and ${\bf D}$ can be written as (\ref{bcond1}) or
(\ref{bcond2}). (The boundary conditions of type ${\bf A}$
coincide with (\ref{bcond1}).) The boundary conditions of type
${\bf B}$ can be written as (\ref{bcond1}) if $ 1+bc \neq 0$
simply by expressing $ \psi'(-0)$ as a function of $ \psi (+0)$
and $ \psi'(+0)$ from the second boundary condition (\ref{bc3}).
If $ 1+bc = 0$ then the boundary condition (\ref{bc3}) is of type
(\ref{bcond2}). The cases $ {\bf C}$ and $ {\bf D}$ can be studied
similarly.

Thus we have proven that these boundary conditions describe all
maximal ${\mathcal P \mathcal T}$-symmetric extensions of $ L_0.$
One can easily check that all these operators are  $ {\mathcal P
\mathcal T}$-self-adjoint as well. The theorem is proven. $ Box $

Similar results in the theory of self-adjoint point interactions
are usually proven using von Neumann extension theory
\cite{book,jmaa,seba}. Instead of developing its counterpart for
${\mathcal P \mathcal T}$-symmetric operators we preferred to give
a constructive proof using boundary conditions.

In what follows the boundary conditions given by (\ref{bcond1})
will be called {\bf connected}, since these conditions connect the
boundary values on the left and right hand sides of the origin of
functions from the domain of the operator. The boundary conditions
of the second type given by (\ref{bcond2}) will be called {\bf
separated}. These definitions and the main proposition are quite
similar to corresponding description of self-adjoint point
interactions given in  \cite{jmaa}. The operators appearing in the
decomposition corresponding to separated boundary conditions are
just the second derivative operators on the half lines with
complex boundary condition of the third type at the origin. Such
non-self-adjoint operators can easily be studied (see e.g.
\cite{pavlov}).

Thus the set of ${\mathcal P \mathcal T}$-self-adjoint point
interactions can be parameterized by $ 4 $ real parameters. The
phase parameter $ \theta $ is redundant in the sense that the
operators corresponding to different values of this parameter are
unitary equivalent.\footnote{Similar fact for self-adjoint point
interactions is described in full details \cite{fei,tomio}.} The
parameterization used in (\ref{bcond1}) and (\ref{bcond2}) is not
optimal in the sense that the correspondence between the
parameters and the boundary conditions is not one-to-one (for
example, if $ b=c=0$, then changing the phases by $ \pi$ we get
the same boundary conditions, similar problem occurs for separated
boundary conditions), but we prefer not to dwell on this point.
 Connected boundary
conditions given by the matrix
\begin{equation}
B = e^{i \theta} \left( \begin{array}{cc} a & b \\
c & a \end{array} \right), \; \; a,b,c \in \Rb, \theta \in [0,
\pi), a^2 -bc = 1
\end{equation}
determine operators which are both self-adjoint and ${\mathcal P
\mathcal T}$-self-adjoint. Separated boundary conditions leading
to such operators are given by formula (\ref{bcond2}) with $
\theta = 0.$

\section{Spectral problems.}

In this section we are going to study the spectrum of second
derivative operators with $ {\mathcal P \mathcal T}$-self-adjoint
point interactions at the origin. Our main result can be
formulated as follows

\begin{theorem}
The spectrum of any ${\mathcal P \mathcal T}$-self-adjoint second
derivative operator with  point interactions at the origin
consists of the branch
 $ [0,\infty)$ of the absolutely continuous spectrum
and at most two (counting multiplicity) eigenvalues, which are
real negative or are (complex) conjugated to each other.
\end{theorem}
{\bf Proof.}
 Let us denote by $A$ any ${\mathcal P \mathcal
T}$-self-adjoint second derivative operator with point interaction
at the origin (described by Theorem 1). To prove the theorem we
are going to calculate its resolvent. Consider the resolvent
equation
$$ (A-\lambda)U = F,~~~~~~F\in L_2(\Rb).$$
The unique solution to this equation is the function $ U$ from the
domain $ {\rm Dom}\, (A)$ of the operator $ A$ satisfying the
differential equation
\begin{equation}
\label{resol} - U'' - \lambda U = F,
\end{equation}
everywhere outside the origin. We denote by $ k$ the square root
of the energy parameter $ \lambda = k^2$, determined uniquely by $
\Im k \geq 0.$ Let us introduce two functions
$$ e_+ (x) = \left\{
\begin{array}{cc}
e^{ikx}, & x > 0 \\
0, & x< 0
\end{array}
\right. , \;\; e_- (x) = \left\{
\begin{array}{cc}
0, & x > 0 \\
e^{-ikx}, & x< 0
\end{array}
\right. .
$$
Then any solution (from $ L_2 (\Rb) $) to (\ref{resol}) can be
written in the form
\begin{equation} \label{re} U (x)= \int_{-\infty}^{\infty} \frac{e^{ik\vert x-y\vert}}{2 ik}
F(y) dy + \rho_+ (F) e_+ (x) + \rho_- (F) e_- (x),\end{equation}
where $ \rho_\pm $ are two parameters to be calculated. The
boundary values of the function $ U$ are
$$
\left\{
\begin{array}{ccl}
U(+0) & = & \displaystyle\frac{1}{2ik} (f_- + f_+) + \rho_+ \\[4mm]
U'(+0) & = &\displaystyle\frac{1}{2} (f_- - f_+) + ik \rho_+\\[4mm]
U(-0) & = & \displaystyle\frac{1}{2ik} (f_- + f_+) + \rho_- \\[4mm]
U'(-0) & = & \displaystyle\frac{1}{2} (f_- - f_+) - ik \rho_-
\end{array}
\right.
$$
where $ f_\pm = \int_{{\Rb}_\pm} F(y) dy .$

Consider the case of separated boundary conditions. Then
(\ref{bcond1}) implies
$$
\left(
\begin{array}{cc}
1 & - e^{i\theta} \left( \sqrt{1+bc} e^{i\phi} - ik b\right)\\[3mm]
ik & - e^{i\theta} \left( c - ik \sqrt{1+bc} e^{-i \phi}\right)
\end{array}\right)
\left( \begin{array}{c}
\rho_+ \\[3mm]
\rho_- \end{array}\right) $$
$$
= \left(
\begin{array}{c}
\frac{1}{2ik} \left( -1 + e^{i\theta} \sqrt{1+bc} e^{i\phi}\right)
(f_- + f_+) +
e^{i\theta} \frac{b}{2} ( f_- - f_+) \\[3mm]
e^{i\theta} \frac{c}{2ik} ( f_- + f_+) + \frac{1}{2} \left( -1 +
e^{i\theta} \sqrt{1+bc} e^{-i\phi}\right) (f_- - f_+)
\end{array}
\right)
$$
This system of equations on $ \rho_\pm$  is solvable if and only
if the determinant of the matrix at the left hand side is
different from zero
\begin{equation}
- e^{i\theta} \left( c - 2ik \cos \phi \sqrt{1+bc} - k^2 b \right)
\neq 0.
\end{equation}
Thus the resolvent equation can be solved for all nonreal $ k$,
which are not solutions to the following quadratic equation
\begin{equation}
\label{disp} b k^2 + 2 i \cos \phi \sqrt{1+bc}\; k - c = 0.
\end{equation}
The two solutions to the last equation
$$ k_{1,2}= - i \frac{\cos \phi \sqrt{1+bc}}{b}  \pm
\frac{\sqrt{(1-\cos^2 \phi) bc - \cos^2 \phi}}{b}$$ are either
pure imaginary or symmetric to each other with respect to the
imaginary axis. The corresponding energy values are real, or
conjugated to each other. Hence the domain $ {\Cb} \setminus (
{\Rb}_+ \cup \{ k_1^2,
 k_2^2 \} )$ belongs to the regularity domain of the operator $ L.$

 Formula (\ref{re}) shows that the difference between the
 resolvents of the operators $ A$ and the unperturbed second derivative
 operator $ L $ has rank two.
The spectrum of the (self-adjoint) operator $ L $ is pure
absolutely continuous and fills in the interval $[0, \infty) $.
The complement to the absolutely continuous spectrum is simply
connected.
 Therefore the perturbed operator $ A$ has
 the branch of absolutely continuous spectrum $ [0,\infty)$ as
 well \cite{naboko,naboko2}.
 Let us study the singularities of the resolvent corresponding to
 the numbers $ k_{1,2}^2.$ Let $\Im k_1 > 0$, then
 the function
 $$
 \psi_1 (x) = \left\{
 \begin{array}{ll}
 e^{i \theta} \left( \sqrt{1+bc} e^{i\phi} - ik_1 b \right) e^{ik_1 x}, & x > 0 ,\\[3mm]
 e^{-ik_1 x}, & x <0;
 \end{array}
 \right.$$
 is a square integrable solution to the equation
 $ -\psi''_1 = k_1^2 \psi_1 , \; x \neq 0 $
 and satisfies the boundary conditions (\ref{bcond1}).
 Therefore this function is a (discrete spectrum) eigenfunction for
 the operator $ A.$ Similarly the function $ \psi_2 = {\mathcal P \mathcal T} \psi_1$
is an eigenfunction corresponding to the eigenvalue $
\overline{k_1}^2.$ If $ k_1$ is pure imaginary ($k_1^2$ negative
real) then the function $ \psi_1$ can be chosen $ {\mathcal P
\mathcal T}$-symmetric or -antisymmetric. In the special case $
(1-\cos^2 \phi)bc - \cos^2 \phi = 0, \cos \phi <0 $ the two
solutions to the dispersion equation (\ref{disp}) coincide.
Nevertheless the corresponding eigenvalue has multiplicity $ 1$,
since the boundary conditions (\ref{bcond1}) are connected.
 Therefore we conclude that every solution to the dispersion
equation (\ref{disp}) from the upper half plane $ \Im k
> 0 $ determines a simple eigenvalue $ k^2$ of the operator
$ A$. Solutions to (\ref{disp}) lying on the nonphysical sheet $
\Im k \leq 0 $ do not determine any eigenvalue of $ L$, since the
corresponding solutions to the differential equation do not belong
to the Hilbert space.
  The theorem is proven for connected boundary conditions.

The proof for separated conditions is quite similar. The only
difference is that in this case the operator $ A$ can be presented
as an orthogonal sum of two operators in $ L_2 ({\Rb}_\pm).$ Each
of these operators can have one (complex) eigenvalue and these
eigenvalues are conjugated to each other. Therefore if the
operator $ A$ has a negative eigenvalue then this eigenvalue has
always multiplicity $2.$ $ \Box $

In particular we have proven the following
\begin{corollary}
The eigenvalues corresponding to
 ${\mathcal P \mathcal T}$-symmetric eigenfunctions of the operator $ A$
are real and negative. Every eigenfunction corresponding to any
real eigenvalue of the operator $ A$ can be chosen $ {\mathcal P
\mathcal T}$-symmetric or -antisymmetric.
\end{corollary}

This proposition holds in fact for any $ \mathcal P \mathcal T
$-self-adjoint operator with non-singular interactions.
 Note that
the spectrum of the operator $ A$ is not always pure real. Let us
study the case when the spectrum is real in more detail. The
spectrum is pure real if and only if the two solutions to equation
(\ref{disp}) are pure imaginary or are situated on the nonphysical
sheet $ \Im k \leq 0 . $ The two solutions are pure imaginary only
if the discriminant is negative $ bc \sin^2 \phi \leq \cos^2 \phi.
$ If the discriminant is positive ($ bc \sin^2 \phi > \cos^2 \phi
$), then the imaginary part of the solutions is given by $ -
\frac{\cos \phi \sqrt{1+bc}}{b} $ and is negative only if $ \cos
\phi > 0. $ We have proven the following
\begin{proposition}
The spectrum of the ${\mathcal P \mathcal T}$-self-adjoint second
derivative operator with connected point interaction at the origin
is pure real if and only if the parameters appearing in
(\ref{bcond1}) satisfy in addition at least one of the following
conditions
\newline
I.
$$ bc \sin^2 \phi  \leq \cos^2 \phi; $$
\newline
II.
$$ bc \sin^2 \phi  \geq \cos^2 \phi \; \; {\rm and}\; \; \cos \phi
\geq 0 . $$
\end{proposition}

 The spectrum does not depend on
the phase parameter $ \theta.$ The operators corresponding to
different values of $\theta $ are unitary equivalent (see
\cite{fei}, where similar results are proven for self-adjoint
point interactions).

\section{Local ${\mathcal P \mathcal T}$-self-adjoint point interactions.}

We consider now ${\mathcal P \mathcal T}$-self-adjoint operators
with point interactions at positions $x = \pm l$, $l \in\Rb_+$.
Every such operator coincides with a certain restriction of the
second derivative operator $ - \frac{d^2}{dx^2} $, defined
originally in $ L_2 ( \Rb)$ on the domain $ W_2^2 (\Rb \setminus
\{\pm l\})$, to the set of functions satisfying some boundary
conditions at the points $ x = \pm l.$ We restrict our
consideration to the case of local connected point interaction
without aiming to describe all possible restrictions leading to
${\mathcal P \mathcal T}$-self-adjoint operators. Suppose that the
functions from the domain of the restricted operator satisfy the
following conditions at $ x= l $
\begin{equation} \label{rel1}
\left(\begin{array}{c}
\Psi (l^+) \\
\Psi' (l^+)
\end{array}\right) = B
 \left(\begin{array}{c}
\Psi (l^-) \\
\Psi' (l^-)
\end{array}\right),
\end{equation}
where the matrix $B = \left( \begin{array}{cc} \alpha & \beta \\
\gamma & \delta \end{array} \right) \in {\rm GL} (2, \Cb)$. Then
to make the operator $ {\mathcal P \mathcal T}$-symmetric we have
to suppose that the following boundary conditions are introduced
at $ x= -l$
\begin{equation} \label{rel2}
\left(\begin{array}{c}
\Psi (-l^-) \\
\Psi' (-l^-)
\end{array}\right) = \left( \begin{array}{cc}
1 & 0 \\
0 & -1 \end{array} \right) \bar{B} \left( \begin{array}{cc}
1 & 0 \\
0 & -1 \end{array} \right)
 \left(\begin{array}{c}
\Psi (-l^+) \\
\Psi' (-l^+)
\end{array}\right),
\end{equation}
The restriction of the second derivative operator to the set of
functions satisfying conditions (\ref{rel1}) and (\ref{rel2}) will
be denoted by $ A $ in this section. Let us study the spectrum of
this operator. The resolvent of the operator $ A $ can be
calculated explicitly using methods of the previous section. One
concludes that the difference between the resolvents of the
operators $ A $ and $L$ has rank four and therefore the operator
$A$ has a branch of absolutely continuous spectrum $ [0, \infty).$
Let us study the discrete spectrum of $ A.$ To calculate the
eigenfunction one can use the following {\it Ansatz}:
\begin{equation}
\psi (x) = \left\{
\begin{array}{ll}
c_1 e^{-ik(x+l)}, & x < - l ; \\
c_2 \cos k (x+l) + c_3 \sin k (x+l), & -l < x < l ; \\
c_4 e^{ik(x-l)}, & x > l;
\end{array}
\right.
\end{equation}
where $ k^2 = \lambda, \Im k > 0 $ and $ c_1, c_2, c_3, c_4 $ are
arbitrary complex parameters to be determined. Substitution of
this function into the boundary conditions (\ref{rel1}) and
(\ref{rel2}) gives the following linear system on $ c_j $
$$ \left(\begin{array}{cccc}
1 & - \overline{\alpha} & \overline{\beta} k & 0 \\
ik & - \overline{\gamma} & \overline{\delta} k & 0 \\
0 & \alpha \cos 2kl - \beta k \sin 2kl & \alpha \sin 2 kl + \beta
k \cos 2kl & -1 \\
0 & \gamma \cos 2kl - \delta k \sin 2kl & \gamma \sin 2kl + \delta
k \cos 2kl & -ik
\end{array} \right)
\left(
\begin{array}{c}
c_1 \\
c_2 \\
c_3 \\
c_4
\end{array}
\right)= 0.
$$
This system has a nontrivial solution if and only if the
determinant of the $4 \times 4$ matrix is equal to zero, which
gives us the dispersion relation
\begin{equation}
\begin{array}{c}
 \displaystyle  \sin 2 kl
\left\{ - k^4 \vert \beta\vert^2 - ik^3 ( \beta \overline{\delta}
+ \overline{\beta} \delta) + k^2 ( \vert \alpha \vert^2 - \vert
\delta \vert ^2) + ik ( \alpha \overline{\gamma} +
\overline{\alpha} \gamma ) - \vert \gamma\vert^2 \right\}
 \\
\\
+ \displaystyle k \cos 2kl \left\{ k^2 (\alpha \overline{\beta} +
\overline{\alpha} \beta) + ik ( \alpha \overline{\delta} +
\overline{\alpha} \delta + \beta \overline{\gamma} +
\overline{\beta} \gamma ) - (\gamma \overline{\delta} +
\overline{\gamma} \delta) \right\}
 = 0.
\end{array}
\end{equation}
Each solution to this equation from the upper half plane
determines an eigenvalue of the operator $ A.$

In what follows we are going to study the operator with the
interaction given by a sum of two delta-functions given by the
following formal expression
\begin{equation}
- \frac{d^2}{dx^2} + (u+iv) \delta(x-l) + (u-iv) \delta (x+l),
\end{equation}
where $ u, v \in \Rb.$ This potential is formally ${\mathcal P
\mathcal T}$-symmetric. This operator is determined by the
boundary conditions chosen as follows\footnote{To define the
operator correctly one can use for example its quadratic form.}
$$ \alpha = \gamma = 1, \; \; \beta = 0 ,\; \; \delta = u+iv. $$
In the  case $ v= 0 $ we get a self-adjoint operator. Let us study
the case $ u= 0, $ when the potential is pure imaginary. The
dispersion relation takes the following simple form
$$ \sin 2kl \left\{ k^2 (1- v^2 ) + 2ik -1
\right\} = 0. $$ All solutions to this equation from the open
upper half plane are determined by the zeroes of the second order
polynomial in the brackets
$$ k_{1,2} = -i \frac{1}{1 \pm v}, \; \; v \neq \pm 1  .$$
These solutions are always pure imaginary. Depending on the
absolute value of $ v $ one or none solutions are situated on the
physical sheet $ \Im k > 0.$ It is interesting to observe that the
spectrum of this operator is always pure real.

\section{Point interactions with real spectrum.}

In this section we are going to discuss connected boundary
conditions at the origin leading to second derivative operators
with real spectrum. Namely the operator $ A = - \frac{d^2}{dx^2} $
with the domain
\begin{equation}
 {\rm Dom} (A) = \left\{ \psi \in W_2^2
(\Rb \setminus \{ 0\}); \left(\begin{array}{c}
\psi (+0) \\
\psi' (+0)
\end{array}\right) = \left( \begin{array}{cc}
\alpha & \beta \\
\gamma & \delta \end{array} \right) \left(\begin{array}{c}
\psi (-0) \\
\psi' (-0)
\end{array}\right) \right\}
\end{equation}
will be studied. We suppose that the matrix $ B = \left(
\begin{array}{cc}
\alpha & \beta \\
\gamma & \delta \end{array} \right) $ appearing in the boundary
conditions is non degenerate (from $GL(2, \Cb)$). Again it is easy
to prove that the operator has branch of absolutely continuous
spectrum $ [0, \infty) $ and it remains to study its discrete
spectrum only. Let us use the following {\it Ansatz} for the
eigenfunction
\begin{equation}
\psi (x) = \left\{ \begin{array}{ll} c_1 e^{-ikx}, & x < 0 \\
c_2 e^{ikx}, & x > 0  \end{array} \right. , \; \; \Im k > 0,
\end{equation}
corresponding to the energy $\lambda = k^2. $ Substituting this
function into the boundary conditions we get the $ 2\times 2$
linear system
$$ \left( \begin{array}{cc}
\alpha - ik \beta & -1 \\
\gamma - ik \delta & - ik \end{array} \right) \left(
\begin{array}{c}
c_1 \\
c_2 \end{array} \right) = 0,
$$
 and the dispersion equation
\begin{equation} \label{di}
k^2 \beta + ik (\alpha + \delta) - \gamma = 0.
\end{equation}
The last equation has the following two solutions
$$ k_{1,2} =
\frac{ - i (\alpha + \delta) \pm \sqrt{- (\alpha+ \delta)^2 + 4
\gamma \beta}}{2 \beta},$$ if $ \beta \neq 0 .$
 Solutions of the
equation situated in the closed lower half plane $ \Im k \leq 0 $
do not determine any eigenfunction of the operator $A.$ Hence the
spectrum of the operator $ A$ is real if and only if both
solutions $ k_{1,2} $ to equation (\ref{di}) satisfy at least one
of the following two conditions:
\newline
1. $ \Im k_{1,2} \leq 0; $
\newline
2. $ \Re k_{1,2} = 0. $

The set of coefficients $ \alpha, \beta, \gamma, \delta $
satisfying the first condition can be parameterized by $ 8$ real
parameters and leads to operators $ A$ with pure absolutely
continuous spectrum $ [0, \infty). $ Pure imaginary solutions to
(\ref{di}) can lead to nontrivial discrete spectrum. Therefore let
us study the set of coefficients satisfying the second condition
in more details.

The solutions $ k_{1,2} $ are pure imaginary if and only if the
following conditions are satisfied:
\begin{equation} \label{coco}
\frac{\alpha + \delta}{\beta} \in \Rb, \; \; \frac{\gamma}{\beta}
\in \Rb, \;\; \frac{4 \gamma}{\beta} \leq \frac{(\alpha +
\delta)^2}{\beta^2}. \end{equation}
 The first two conditions imply
that the complex numbers $ \tau = \alpha + \delta, \beta,$ and $
\gamma $ have the same phase. Therefore let us introduce the
following parameterization:
$$ \tau = t e^{i\theta}, \; \; \beta = b e^{i\theta},
\; \; \gamma = c e^{i \theta}, $$ where $ t,b,c $ are real
numbers. In order to guarantee that solutions $ k_{1,2} $ are pure
imaginary the real parameters should satisfy the inequality
\begin{equation}
4 \frac{c}{b} \leq \frac{t^2}{b^2}.
\end{equation}
We conclude that the second family can be parameterized by $ 6$
real parameters. The intersection between the two families of
parameters leading to operators with pure real spectrum is not
empty, but the second family is not included in the first one.

The four-parameter family of self-adjoint (connected) boundary
conditions
 is contained in the second ($6$-parameter) family just described.
Parameters leading to ${\mathcal P \mathcal T}$-self-adjoint
boundary satisfy the first two conditions (\ref{coco}). Hence the
family of ${\mathcal P \mathcal T}$-self-adjoint (connected)
boundary conditions leading to operators with real spectrum
belongs to the second family of boundary conditions as well.

\section{Conclusions.}

The relations between ${\mathcal P \mathcal T}$-self-adjoint,
self-adjoint and real second derivative operators with point
interactions have been studied. One has to investigate whether the
operators with real spectrum are similar to certain self-adjoint
operators or not. It is easy to generalize these results in order
to include point interactions with the support at arbitrary finite
(or even infinite) number of points. The results obtained can be
applied to construct exactly solvable few-body systems with
unusual symmetries (following \cite{spin1,rmphys}). This method
can be generalized to include point interactions in spaces of
higher dimension and higher order differential operators (like in
\cite{boman}).


\end{document}